# Large-scale clustering amongst Fermi blazars; evidence for axis alignments?

M.J.M. Marchã[1] * & I.W.A. Browne[2]
[1] *University College London, Department of Physics and Astronomy, Gower Street, London, WC1E 6BT, UK*
[2] *Jodrell Bank Centre for Astrophysics, Department of Physics and Astronomy, University of Manchester, M13 9PL, U.K.*

**ABSTRACT**
We find evidence for large-scale clustering amongst Fermi-selected BL Lac objects but not amongst Fermi-selected FSRQs. Using two-point correlation functions we have investigated the clustering properties of different classes of objects from the Fermi LAT 4FGL catalogue. We wanted to test the idea based on optical polarization observations that there might be large volumes of space in which AGN axes are aligned. To do this we needed a clean sample of blazars as these are objects with their jet axes pointing towards the observer and Fermi sources provide such a sample. We find that high latitude Fermi sources taken as a whole show a significant clustering signal on scales up to 30 degrees. To investigate if all blazars behave in the same way we used the machine learning classifications of Kovačević, et al. (2020), which are based only on gamma-ray information, to separate BL Lac-like objects from FSRQ-like objects. A possible explanation for the clustering signal we find amongst the BL Lac-like objects is that there are indeed large volumes of space in which AGN axes are aligned. This signal might be washed out in FSRQs since they occupy a much larger volume of space. Thus our results support the idea that large scale polarization alignments could originate from coherent alignments of AGN axes. We speculate that these axis alignments may be related to the well-known intrinsic alignments of galaxy optical position angles.

## 1 INTRODUCTION

In the widely accepted concordance model of cosmology the largest structures expected in the Universe are ≤1 Gpc (Yadav et al., 2010; Marinello et al., 2016). However, there have been observational claims for the detection of significantly larger structures, for example amongst the polarization alignments of quasars (Hutsémekers, et al., 1998, 2005, 2011) and amongst GRBs (Horváth et al., 2014, Horváth et al., 2015, Horváth et al. 2020, Balazs et al., 2015).Since these claims, if substantiated, would have profound implications for large scale structure formation it is important to see if it is possible to find independent evidence to back them up.

One of the difficulties encountered when looking for evidence of very large structures is the uniformity of available surveys over large areas. Optical surveys have to contend with the effects of patchy Galactic extinction and radio surveys often have subtle variations in sensitivity over different parts of the sky that might mimic large scale clustering. Possibly the cleanest sample of extragalactic sources is that produced by the LAT instrument on the Fermi gamma-ray satellite (Atwood et al., 2009). The vast majority of high latitude sources detected by Fermi are blazars; i.e. a combination of BL Lacs objects and flat-spectrum radio quasars (FSRQs). In the 4FGL (v19 - https:.gsfc.nasa.gov/ssc/data/access/lat/8yr_catalog/) "clean sample" ( e.g sources without analysis flags), there are 2649 objects (Fermi collaboration, 2020) and these objects are subdivided according to associations with existing catalogued objects. There are 1018 sources listed as BL Lacs, 598 as FSRQs and 972 as blazar candidates. In the past Allevato et al. (2014) analysed the projected correlation function of 2FGL Fermi blazars on scales of degrees and concluded on the basis of the similarity of their clustering properties that blazars occupy massive dark matter haloes. They also conclude that BL Lacs and FSRQs are objects residing in similar dense environments. They did not report any results concerning clustering on scales ≥6 degrees. Others have examined the clustering of other types of AGN; for example, Goncalves et al. (2020) have looked at the distribution of SDSS quasars while Charutha et al.(2020) have looked at X-ray selected AGN. Goncalves et al. find the distribution of high redshift quasars to be consistent with the ΛCDM predictions. Charutha et al. find that their X-ray selected objects have clustering properties typical of galaxies in the mass range $10^{12} - 10^{13}$ solar masses.

We have studied the clustering properties of Fermi blazars but on larger angular scales than Allevato et al. As stated above we were motivated by the claim based upon various observations, mostly but not exclusively in optical polarization, that there are large scale alignments and anisotropies amongst active galaxies, namely:

• The claim that the optical polarization position angles measured for quasars have coherence over regions of tens of degrees (Hutsémekers, 1998; Hutsémekers et al.,2005; Hutsémekers et al., 2011) but see also Tiwari & Jain (2019) and Shurtleff (2014) for further discussion on alignments for different samples.

• The existence of two large quasar groups (LQGs) of 450 Mpc

* e-mail:m.marcha@ucl.ac.uk

in size (Clowes et al., 2013). It, however, should be noted that Park et al., (2015) challenge the statistical significance of the largest LQGs

• The correlation of LQG elongations with radio polarization position angles (Hutsémekers et al., 2014; Pelgrims & Hutsémekers, 2015, Pelgrims & Hutsémekers, 2016).

• The clustering of radio source axes on scales of 6 degree in the LoTSS survey (Osinga et al. 2020, arXiv:2008.10947)

• The claim that gamma ray bursts are clustered on scales of tens of degrees (Horváth et al. 2015, Horváth et al. 2014, Horváth et al. 2020, ).

• Evidence for an unexpectedly strong dipole in the distribution of radio sources found in large area surveys like NVSS, SUMSS, WENSS and TGSS (Siewert et al., 2020, and references therein). Also a dipole stronger than expected is also found when the distribution of quasars detected in the infrared by WISE is analysed (Secrest et al, 2020) and for AGN in the infrared-selected MIRAGN sample (Singal, 2021).

If these claims, particularly those related to the polarization alignment, are substantiated the most likely astrophysical interpretation of the polarization position angle results is that the axes, presumably the angular momentum axes of the black holes powering the AGN, are coherently aligned over large volumes of space. It is important to our investigation to note that the alignment of AGN axes with respect to the line of sight to the observer is what results in a radio source being seen as core-dominated and/or being classified as a blazar according to unified schemes of radio sources (e.g. Orr & Browne 1982, Antonucci & Ulvestad 1985, Urry & Padovani 1995). Thus a prediction based on the idea that AGN axes align on large scales is that blazars, since they are objects that almost certainly have jet axes that point close to the line of sight to the observer, should show an apparent large scale clustering signal.The exact angular scale over which one might expect to see alignments is uncertain. But it will be larger than the expected $\sim 1/\gamma$ beaming angle (where $\gamma$ is the Lorentz factor) required for an object to be recognized as a blazar simply because in any volume of space there will be a distribution in the alignment angles of AGN. Also we do not know what physical scale over which the alignments of AGN axes might occur. For these reasons we look for a clustering signal over a range of angular sizes. We emphasize that any clustering signal that we might find does not necessarily represent a variation in the space density of objects but only a coherent grouping of their axes alignment angles.

Not only do Fermi-selected sources represent one of the cleanest all-sky sample of extragalactic sources, because they are nearly all blazars, they also represent a particularly suitable sample to look for possible indirect consequences of large scale axes alignments. Thus based on the above chain of reasoning we have looked to see if there is a clustering signal amongst Fermi 4FGL blazars believing them to be less susceptible to selection effects than other samples of potentially aligned objects. Our hypothesis is that there should exist patches of sky with high concentration of blazars and other patches, where the axes are not aligned, in which there is a deficit. In practice, because the range of solid angles for which an object appears as a blazar is roughly two orders of magnitude smaller than when it will not, the blazar clustering should be much more prominent than in the non-blazar parent sample. This assumes that the delineation between blazar and non-blazar occurs at an angle to the line of sight $\sim 10$ degrees.

We are not the first to investigate alignments of the orientations of radio source axes motivated by the quasar optical polarizations results. Contigiani et al. (2017), Panwar et al (2020) and Osinga et al. (2020) have adopted a direct approach and have examined the distribution of structural position angles of radio sources in the FIRST survey (Panwar et al.), FIRST and TGSS (Contigiani et al.) and LoTSS survey (Osinga et al.). Both investigations find some evidence for alignments. Contigiani et al. find that there is a ~2% probability that the alignments they see on scales of 1.5 degrees could arise by chance while Osinga et al. find evidence for position angle alignments in a two dimensional analysis on a scale of ~4 degrees and with a formal probability of arising by chance of $\leq 10^{-5}$. However, they also find that this result does not hold up in their three dimensional analysis. The authors suggest that there is some unknown bias in the derived LoTSS survey parameters which might account for the apparently very significant two dimensional detection.

The current paper is organized as follows; we first describe our two-point clustering analysis and the tests we have performed. We then calculate the two-point correlation function for 4FGL objects. Many BL Lac objects do not have redshifts so for our analysis we ignore redshift information for all objects. We start with the analysis of the 4FGL sample as a whole and then divide according to the associations listed in the catalogue, namely, BL Lacs, FSRQs and blazar candidates of unknown type (BCUs). Since in our initial analysis we find that the objects associated with BL Lacs and FSRQs have different clustering properties we suspected that this could arise from a selection effect, particularly affecting BL Lac associations. Therefore, in order to avoid any uncertainties that might arise from using catalogued associations which depend on the availability of optical data, we supplement the 4FGL associations with the machine learning classification of BCUs by Kovačević, et al., (2020) which are based on gamma-ray information only. Henceforward we use the terminology BL Lac-like (and FSRQ-like) to refer to the ensemble of sources with BL Lac (and FSSQ) properties, ie, the sum of known associations and those BGUs classified as one or the other by Kovačević et al. (2020). Finally, we will discuss if the difference we find between the clustering properties of BL Lac-like and FSRQ-like objects is of astrophysical origin.

## 2 THE TWO-POINT ANALYSIS

Following others who examined the clustering properties of sources in radio surveys (e.g. Blake & Wall (2002); Blake et al. (2004); Overzier et al. (2003); Rana & Bagla (2018) we use the Landy & Szalay (1993) method to derive the angular two point correlation functions. We emphasize that often we are primarily interested in the relative behaviour of different sub-samples of objects, in which case, the precise analysis technique applied to the different sub-samples selected from the same survey is of secondary importance and should not affect our conclusions. This is also true when we compare the observational results with simulated catalogues, provided we ensure that the sky coverage is identical. However, as an initial cross-check of our software we have repeated the analysis of the distribution of NVSS sources by Overzier et al., both with and without masking. We get essentially the same results as others for both versions of the analysis, e.g we recover the same overall fit to the data within the errors. We have also checked that when we analyse artificial samples of randomly distributed objects we detect no clustering signal.

The basic idea behind the two-point correlation function is to measure the excess (or deficit) probability of finding a pair of sources separated by a certain distance. This excess (or deficit) probability is measured by comparing the measured separation of two sources with that expected for another two sources that are scattered randomly on the sky. In practical terms what needs to be done in order to estimate the two-point correlation function for a given survey (D) is to count the number of pairs of sources as a function of separation (DD) and then to divide that number by the number of pairs that would be expected for an 'un-clustered' (random-R) distribution. Hence, for a given catalogue, one requires to: (1) construct a random catalogue with an identical footprint to that of the real data, and large enough not to introduce significant Poisson errors (Coil012), and (2) adopt an approach to estimate the ratio between data pairs (DD) and the data-random (DR) and random-random (RR). Though different approaches have been used to estimate this ratio, the one most commonly used, and the one that will be used in this work is that of Landy & Szalay (1993).

The routine used throughout this work to estimate the two point angular correlation function is the *bootstrap two point angular* algorithm which is discussed in the AstroML python package (see VanderPlas, J. T. et al. 2012, and Ivezić, Z. et al., 2019). As mentioned before, the random catalogue within the routine has to be matched to the sample in question. In particular, the two_point_angular routine makes use of an algorithm ( *uniform_sphere*) to do just this, e.g, it enables one to distribute a number of sources over the entire sky in un-clustered way. It is upon this 'uniform sphere' that the masks relevant to the survey are then applied. Since the 4FGL is essentially an all-sky survey the only masks used were those to match any limit imposed on the galactic latitude $b$. With a limiting $|b| > 20$ degrees we use roughly 660000 sources in the random catalogue. This a much larger number than those in the typical datasets which we analyse which contain a maximum of roughly 2200 objects.

The error bars plotted on the individual points of the two-point angular correlation function were estimated via the bootstrap technique (Efron, 1978, but see Norberg 2009, for an expanded review on errors in clustering measurements). Essentially this technique involves N realizations of the random catalogue for which the value and the error on each point of the two-point correlation function correspond to the mean and standard deviation of the N values obtained, respectively. There is however another (somewhat hidden) parameter in this bootstrap technique which corresponds to the number of random points ($n_{rand}$) used in the random catalogue in relation to the number of data points ($n_{dat}$). The original routine (*bootstrap two point angular*) uses $n_{rand} = 2n_{dat}$. As discussed in Norberg (2009), the size of the errors is influenced by the choice of these two numbers (N and $n_{rand}$). Having experimented with both, it was found that a good compromise between error bars and computing time was to use N=100 and $n_{rand} = 4n_{dat}$ for each of the N bootstraps (which incidentally is in agreement with the findings of Norberg (2009).

The values of the individual points in the correlation function are not independent and the error bars shown in the plots are not used to derive the statistical significance of any clustering. Instead the method used to assess the significance of the result is outlined in Section 3.1.

For the analysis of the 4FGL sample we examine the two point correlation function on angular scales up to 30 degrees. This choice for the range of angular scales to be investigated is partially motivated by the previous claims suggesting there might be alignments of AGN optical polarizations on similar scales. For our analysis we exclude low galactic latitude objects. We initially excluded sources with $|b|$ <10 degrees but in the end excluded all objects with $|b|$ <20 degrees. We adopted this ultra conservative approach in order to ensure that there is no contamination of the sample from Galactic sources and that any sensitivity variation of Fermi LAT that arises from Galactic emission does not affect out results.

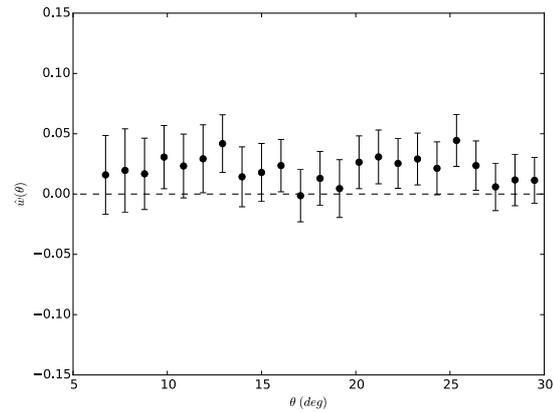

**Figure 1.** The two-point angular correlation function in the range 6 to 30 degrees for all blazars in the 4FGL clean sample for $|b|$ >20 degrees.

## 3 ANALYSIS OF THE 4FGL SAMPLE

In Figure 1 we show the two-point correlation function for the whole 4FGL Clean Sample for $|b| > 20$ excluding only the 53 objects which have non-blazar associations. Most of these 53 excluded objects are galaxies. Since we have imposed Galactic latitude cuts, virtually all the sources will be blazars. As stated above, we excluded sources with $|b| < 20$ degrees because the clustering signal is slightly stronger when low latitudes are included and this may be due to Galactic emission and Galactic sources contaminating the sample.

The first result is that there is significant clustering signal for all blazars (the sum of BL Lacs, FSRQs and BCUs) on scales of 6 degrees and greater. This is clear from the excess of positive values in Figure 1 (see also Table 1 and the discussion below for an estimate of the statistical significance). We focus on scales greater than 6 degrees because we are looking for possible pseudo-clustering effects arising from axis alignments rather than those arising from real density fluctuations which dominate at smaller separations (Blake & Wall 2002, Blake et al. 2004, Goncalves et al. (2020))

A strong motivation for using Fermi blazars as the sample to investigate is that they represent a clean all-sky sample free from any selection effects that might give rise to a spurious clustering signal. Given that there is a strong clustering signal for blazars as a whole it is interesting to see if different types of object show the same behaviour. In the 4FGL catalogue objects are divided into BL Lacs, FSRQs and BCUs (blazar candidates of uncertain type). However, to be classified as either a BL Lac or an FSRQ an association with an optical object has to be made. This highlights a potential problem. If we divide the 4FGL sample into BL Lacs and FSRQs based on *associations* listed in 4FGL we no longer have a sample free from selection effects. This is a significant concern because the probability of finding an association depends on the availability of external data like optical spectroscopy. The availability of such data, however, is likely to vary significantly from one part of the sky to another, something that will affect the BL Lac population more severely than FSRQs as for the former optical confirmation of their nature requires higher signal to noise optical spectroscopy to establish the absence of strong emission lines than for FSRQs (which have strong emission lines). Though the listed associations are reliable they as not complete and this may result in a spurious clustering signal when we use them. Indeed, we have found that, if we use our two point analysis on BL Lac associations, we get a suspiciously strong clustering signal.

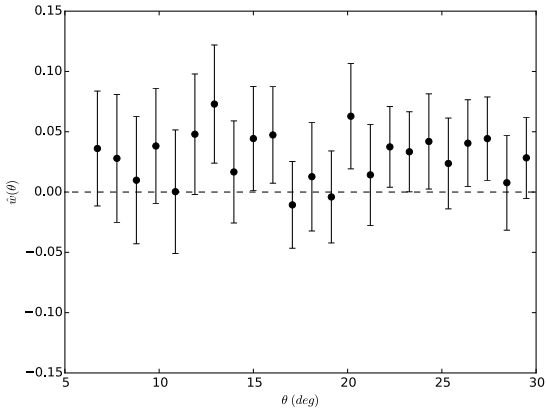

**Figure 2.** The two-point angular correlation function for the BL Lac-like for $|b| > 20$ degrees.

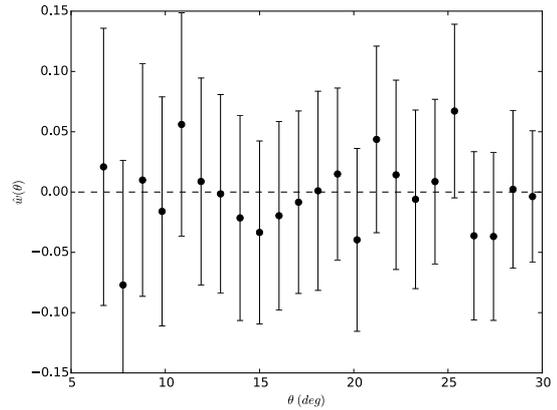

**Figure 3.** The two-point angular correlation function for FSRQ-like for $|b| > 20$ degrees.

For the above reasons we decided to make use of the machine learning classifications for 4FGL BCUs listed in Kovačević et al. 2020. We note that Kang et al., 2019 have also performed a machine learning analysis on 4FGL. However, for simplicity we will report our results only obtained using the Kovačević et al. classifications though we obtained similar results using the Kang et al. classifications.

The advantage of machine learning classifications is that they are produced by algorithms and are based entirely upon the catalogued gamma-ray properties of the sources. Therefore the BL Lac-like and FSRQ-like sub-samples should be almost free from any external biases as to where sources are located being that they treat all parts of the sky equally. There is a possibility that some residual bias might remain if the machine learning classifications are very unreliable since miss-classifications of BL Lacs as FSRQs might still leave a small deficit of BL Lacs in the areas in which there is an existing deficit of optical spectroscopic data required to give a BL Lac association. Three things reassure us that this is not a serious problem: (i) the ratio of machine learning BL Lac to FSRQs is the same within the errors as that found for the associations, (ii) the machine learning classifications are claimed to be more than 98% reliable for the majority of BCUs, and (iii) those objects without 98% reliable classifications do not themselves show any residual clustering signal (See Table 1). The results using the combination of known associations and machine learning classifications are listed in Table 1.

The key finding is that there a significant difference in the clustering signal for the BL Lac-like and FSRQ-like objects. Whereas the BL Lac-like objects show a strong clustering signal (7.5$\sigma$), the FSRQ-like objects behave like a randomly distributed sample (See Figure 2 and Figure 3).

### 3.1 Assessing the statistical significance

It is essential to be able to establish whether or not these apparently different clustering signatures for BL Lac-like and FSRQ-like objects are statistically significant. First we note that the clustering signals for blazars as whole, and for BL Lac-like objects separately, appear to be present at approximately the same level on all angular scales $\geq 6$ degrees. With this in mind a straightforward test is to simulate 1000 random distributions for each class of object and see how often the random distributions produce a clustering signal greater than or equal to that observed in the real sample. We do this by using the routine 'uniform_sphere' introduced previously to create a un-clustered distribution of 1 million sources over the entire sky. Just as before, the mask used for the survey is applied to this 'un-clustered sky' and from which 1000 random distributions are drawn for each data sample (e.g $N_{sample}$ equal to 2153 for Blazars, 1220 for BL Lac-like, and so on). For each of these randomly drawn samples we ran the same procedure to estimate the two-point angular correlation function and the parameters derived from it, just as we did for the real data.[1]

| Type | Number | Mean | Probability ($\sigma$) |
|---|---|---|---|
| All Blazars (including BCUs) | 2153 | 0.021 | 8.9 |
| BL Lacs & FSRQ | 1386 | 0.070 | 20.1 |
| BL Lac-like | 1220 | 0.029 | 7.5 |
| FSRQ-like | 616 | -0.002 | 0.1 |
| BL Lac-like & FSRQ-like | 1836 | 0.026 | 9.8 |
| Unclassified BCU | 317 | -0.004 | 0.2 |

**Table 1.** Table 1. The numbers of 4FGL 'Complete Sample' objects for $|b| > 20$ in different classes *columns 1 and 2), the mean value of the two-point correlation function on scales from 6 to 30 degrees (column 3) and the probability in terms of $\sigma$, that the mean value would arise by chance in a randomly distributed sample (column 4).

Having found the means of the correlation function values for each of the 1000 random runs corresponding to each of the data samples, a histogram of these means can be created. The histograms are approximately Gaussian hence enabling the estimate of the standard deviation of the distribution, as well as of the probability that a value greater than or equal to that of the observed signal would be found by chance.

In Figure 4 we show an example histogram for a simulated random sample containing 2153 objects which matches the size of all the 4FGL blazars. The observed value for the mean is marked by a vertical line and in this case the unambiguous conclusion is that the sources are clustered with a high significance level (and a probability of occurring by chance). We also show the histograms for BL Lac-like objects (Figure 5) and FSRQ-like objects (Figure 6).

---

[1] We also did a check to see if there was a significant difference between drawing 1000 random samples from one 'un-clustered universe', or by creating 1000 'un-clustered universes' from which the $N_{sample}$ draws were made. No difference was found.

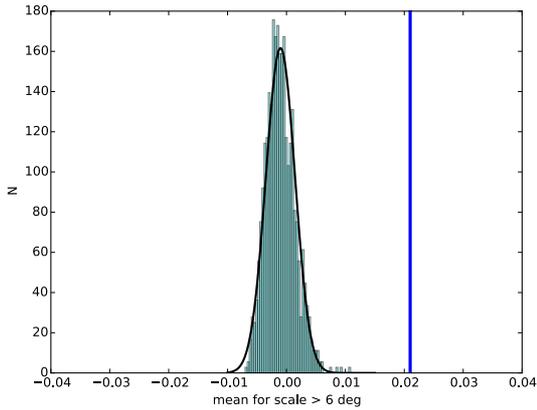

**Figure 4.** The histogram of average values of the two point angular correlation fucntion for scales between 6 and 30 degrees obtained from 1000 simulated random samples containing 2153 objects which matches the size of all the blazars (excluding the 53 non-blazar associations) in the 4FGL clean sample and $|b| > 20$. The blue vertical line shows the mean value (between 6 and 30 degree scales) for the real sample.

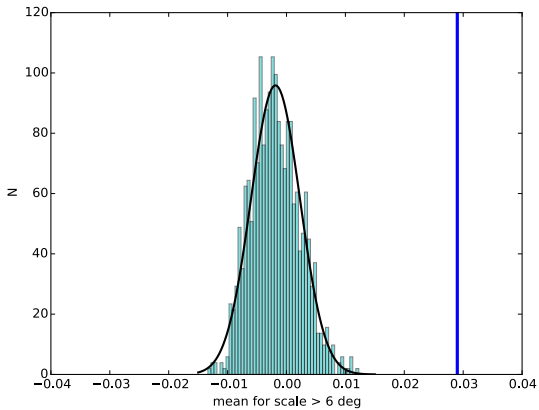

**Figure 5.** The histogram of average values of the two point angular correlation fucntion for scales between 6 and 30 degrees obtained from 1000 simulated random samples containing 1220 objects which matches the size of the BL Lac-like sample ($|b| > 20$). The blue vertical line show the mean value (for scales between 6 and 30 degrees) for the real sample.

## 4 DISCUSSION

The important results are that we have found a significant clustering signal for all the blazars (BL Lacs + FSRQ + BCUs) and that, when divided on the basis of machine learning results into BL Lac-like objects and FSRQ-like objects, these sub-samples behave differently, with the former showing strong clustering.

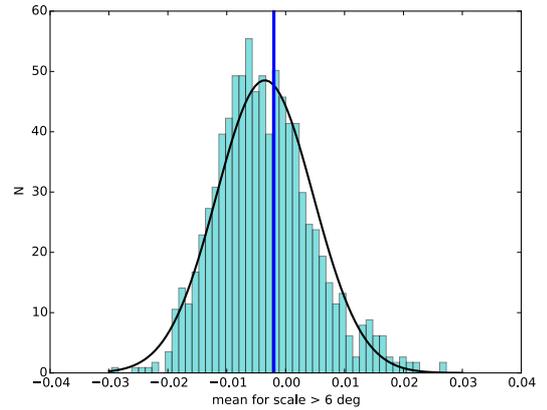

**Figure 6.** The histogram of average values of the two point angular correlation fucntion for scales between 6 and 30 degrees obtained from 1000 simulated random samples containing 616 objects which matches the size the FSRQ-like sample ($|b| > 20$). The blue vertical line show the mean value (for scales between 6 and 30 degrees) for the real sample.

### 4.1 Evidence for clustering using machine learning classifications

When the objects classified by machine learning algorithms as listed by Kovačević et al. are considered, a significant clustering signal on scales in excess of 6 degrees is present for the objects classified as BL Lac-like but not for those classified as FSRQs. There are two things to note about the strength of the clustering signal for BL Lac-like objects. The first is that it is much stronger than that seen in other samples, for instance amongst the radio-loud objects studied by Blake et al. (2003). In their analysis of radio sources from the SUMMS and NVSS surveys they find a value for $\omega(\theta)$ $\sim 10^{-3}$ at an angle of 10 degrees. We find $\omega(\theta)$ to be 0.029 on this same angular scale. The second remarkable thing is that we find no clear evidence for a falloff in the strength of the clustering signal on scales ≥10 degrees as predicted in ΛCDM models. For example, in Figure 6 of Tiwari et al. (2021) the expected $\omega(\theta)$ distribution is shown over plotted with the observed distribution for radio sources in the LoTTS survey. Both show the two-point function falling steeply with increasing angle. Therefore our results are both quantitatively and qualitatively different to what is seen and expected for *normal* cosmological clustering, suggesting a different origin.

Our results are difficult to explain away as arising from some selection effect. The takeaway message is that formal probability of the observed clustering for BL Lac-like objects with Galactic latitudes > 20 degrees arising by chance is $\leq \times 10^{-15}$, while at the same time the FSRQ-like show no evidence for large scale clustering. This pronounced difference in clustering signal between the FSRQs and the BL Lac-like objects gives us real confidence that when do see a signal, as for BL Lac-like objects, it should be believed. It is hard to see how it can be a result of some survey selection effect, or some shortcoming in our analysis. All objects come from the same survey, the same area, and have been analysed in an identical way. Our conclusion is that gamma-ray selected BL Lac objects display apparent clustering on scales up to 30 degrees.

A further somewhat more tentative conclusion is that the clustering signal seen for blazars as a whole is likely to be virtually all contributed by the BL Lac fraction of the blazar sample. We do, however, find it slightly surprising that the combined signal for all blazars, and for the sum of BL Lac-like and FSRQ-like objects, is more statistically significant than for just for BL Lac-like objects (see Table 1). Why should adding in an unclustered set of objects enhance the statistical significance of the signal? But we note that, though the statistical significance of the combined signal for blazars and the sum of the BL Lac-like and the FSRQ-like objects is higher than for BL Lac-like objects, the clustering amplitude of the blazars and blazar-like objects is less than that for the BL Lac-like (0.021, 0.026 and 0.029 respectively). Hence, by this measure, adding in the FSRQs has reduced the strength of the signal as expected.

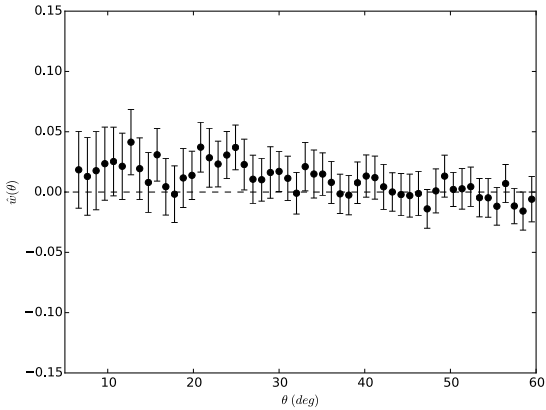

**Figure 7.** The two-point correlation function extending up to 60 degrees for all blazars in the 4FGL Clean Sample with $|b| > 20$.

### 4.2 The clustering scale

To keep things simple we have focused on scales up to 30 degrees but there are clear indications that the strength of clustering decreases for scales ≥30 degrees. This is illustrated in Figure 7 which shows the two point correlation function for the full blazar sample extending up to 60 degrees where it is apparent that any clustering signal falls to close to zero at the maximum angular scale. The mean value for the function in the range 30 to 60 degrees is 0.002 compared to 0.021 for the range 6 to 30 degrees. Assuming for the moment that the clustering scale is ~20 degrees for the BL Lac-like, how consistent is this with the widely accepted concordance model of cosmology in which the largest structures expected in the Universe are ≤1 Gpc (Yadaf et al., 2009; Marinello et al., 2016)? The typical redshifts for the different classes of object are relevant. We show the redshift distributions for associated FSRQs and associated BL Lacs in Figure 8. The median redshift for BL Lacs is 0.344 and for FSRQs is 1.15. It should of course be noted that many BL Lacs do not have measured redshifts and thus the true median redshift for BL Lac-like objects is likely to be somewhat larger. If we assume that BL Lacs have typical redshifts of 0.35, clustering on a scale 20 degrees corresponds to a linear scale of ~350 Mpc which does not violate the concordance model constraint. On the other hand for quasars of redshifts ~1.1, 20 degrees corresponds to 600 Mpc. In terms of comoving volume, quasars occupy ~ 20 times that of BL Lac objects. Thus the signal from individual "clustering cells" of scale ~350 Mpc would be smeared out. Therefore, if we had found a clustering signal amongst quasars this would begin to challenge orthodoxy. Thus it should not be too surprising that the FSRQs show little signal on the scales we have investigated here.

### 4.3 Interpretation in terms of axes alignments.

We set out to test the idea that the polarization results reported by Hutsémekers, (1998); Hutsémekers et al.,(2005); Hutsémekers et al., (2011) could be a result of large scale alignments of AGN axes in which case it might manifest itself in the clustering properties of blazars. We have found strong evidence that there is a clustering signal for Fermi blazars and the indications are that this clustering is confined to the dominant subset of Fermi blazars, the BL Lacs. We regard this as support for the idea that there are coherent axis alignments amongst AGN in general on scales approaching 0.5 Gpc.

Are there alternative explanations for our results? The most obvious would be that there is real clustering of BL Lac objects on scales of tens of degrees and what we see has nothing to do with axis alignments. We reject this possibility because, if true for BL Lac objects, similar clustering should also be found amongst the class of object that host BL Lacs, i.e. luminous red galaxies No such clustering of the strength we see is found on the necessary scales (e.g, Sawangwit et al., 2011). Furthermore, other types of AGN do not display similar clustering on these scales. For example, Goncalves et al. (2020) have looked at the distribution of SDSS quasars while Charutha et al.(2020) have looked at X-ray selected AGN. There is one caveat. The lack of evidence for large-scale clustering is of course only true provided one discounts the anomalous strength of the dipole reported amongst radio sources (Siewert et al., 2020), amongst quasars by Secrest et al. (2020) and amongst infrared-selected AGN (Singal 2021) as evidence for clustering.

### 4.4 Independent evidence for alignments amongst extragalactic objects.

Our results imply that, for at least a subset of radio-loud objects, there are large areas of sky over which radio source axes align. Is there any evidence for this when the structures of radio sources are examined directly? Taylor & Jagannathan (2016) and Contigiani et al, (2017) have found evidence for a non-uniformity of radio source position angles over large regions of sky. More recently Osinga et al. have studied the alignments of radio sources axes amongst sources found in the LOFAR 150 MHz survey, LoTTS. At present this survey covers the relatively small area of 424 square degrees but will ultimately cover the whole northern sky. They find somewhat conflicting evidence some of which suggests the existence of coherent alignments on scales of ~4 degree. Here we note that, because of the small area covered, Osinga et al. are not able to see alignments on scales of 10s of degrees that our analysis has revealed. It will, however, be interesting to see the results when the full LOFAR survey becomes available for analysis.

It has been known for a long time that the optical axes of galaxies on scales of clusters and filaments have a tendency to align (Schneider & Bridle, 2010, MNRAS; Joachimi et al., 2015); these are known as intrinsic alignments and are of great concern since they can mimic the effects of weak gravitational lensing. Physically, the alignment is believed to be produced by tidal torques (Chisari, et al. 2015). Recent investigations of intrinsic alignments have focussed on them as a potentially useful cosmological probe (Okumuta et al., 2019; Yao et al., 2020; Kurita et al., 2020). Desai & Ryden (2020) in their analysis of SDSS galaxies find evidence for alignments amongst early type galaxies on scales up to 30 Mpc. Of relevance to our work, Yao at al. find that the alignment effect has a strong dependence on colour getting stronger as galaxies get redder. Simulations indicate that alignments should be detectable on linear scales large enough to see baryon acoustic oscillations; i.e. ~ 100 Mpc (Kurita et al, 2020).

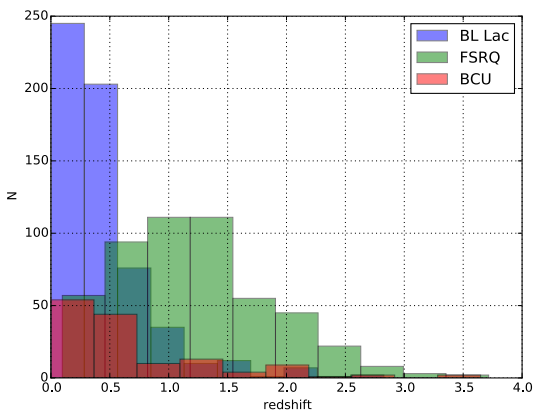

**Figure 8.** The histogram of the redshifts of 4FGL Clean Sample $|b| > 20$ BL Lacs, FSRQs and blazars of uncertain type (BCUs).

For intrinsic alignments to be related to what we see amongst BL Lac objects, it is clear that there would have to be a correlation between the optical elongation of host galaxies and the radio source axes. Battye and Browne (2009) found that in red, early-type, galaxies there was clear evidence for an alignment of radio axes with the minor axes of their optical hosts. Importantly this is strongest amongst low radio luminosity objects. BL Lacs are usually hosted by optically luminous red galaxies and furthermore the intrinsic radio luminosities of BL Lacs are low as indicated by their extended emission. Thus it is likely that the required correlation between optical axes and radio axes exists for BL Lac hosts and therefore we suggest that it is entirely conceivable that what we are seeing in terms of BL Lac clustering is another, larger scale, manifestation of the alignment effect.

### 4.5 Other possible manifestations of coherent alignments of AGN axes.

The evidence we have presented suggests that on scales ∼350 Mpc there are coherent alignments of blazar axes and, by extension, AGN axes in general. If this is true there should be other manifestations of the same phenomenon and in this section we give a brief critique of possible new investigations. One of the most obvious is to extend the work like that of Osinga et al. (2020) who have looked for alignments of the axes of extended sources found in radio surveys. New surveys with many more extended radio sources in very large areas of sky are becoming available like EMU with ASKAP (Norris,2011) and we look forward to the analysis of these. The advantage of this approach is that it is direct and, unlike the approach we adopt, does not require any prior selection to identify a particular class of object; i.e. blazars. Another fairly direct approach would be to gather more measurements of optical and radio polarization position angles in order to extend the work of Hutsemekers et al. and Pelgrims et al.

There are other possibilities using presently available data. One which we are pursuing (Marchã & Browne in preparation) is to use existing large area radio surveys to select out sources of different radio spectral types. The expectation is that flat spectrum radio sources have their axes pointing at us and therefore should show a clustering signal whereas steep spectrum sources from the same surveys should be essentially un-clustered. The challenge in such investigations is to ensure uniformity of selection in the face of unknown systematic effects in the radio surveys used.

## 5 SUMMARY AND CONCLUSIONS

We set out to investigate the clustering properties of blazars on scales of 10s of degrees in order to see if the polarization position alignment results of Hutsemékers and collaborators could be a sign of a wider phenomenon. We predicted that there might be an apparent clustering of blazars based on the idea that blazars will only be seen as such when their axes point towards the observer. The sample of blazars we chose to investigate was taken from the 4FGL survey made with the LAT instrument on the Fermi gamma-ray satellite. The survey covers the whole sky and the high latitude sources in 4FGL are almost exclusively blazars and they represent one of the cleanest blazar samples available. We find a surprisingly strong clustering amongst the high latitude Fermi 4FGL blazars. To investigate further we split the blazars into sub-classes. A problem with doing this is that, if we rely on the listed associations of gamma-ray sources with optical objects, there is a danger that we introduce position-dependent bias. Hence, in order to preserve the desirable properties of the sample when splitting into different sub-classes we supplemented the known associations with those BCUs classified as either BL Lacs or FSRQs by Kovačević, et al. (2020) using machine learning. Thus the BL Lac-like and FSRQ-like sub-samples are almost entirely based on gamma-ray properties alone and therefore they should be free from any potential biases associated with identifying the gamma-ray sources with optical counterparts.

Using our two-point correlation function analysis we find highly significant clustering of the BL Lac-like objects on scales of up to 30 degrees. The signature of this clustering is both quantitatively and qualitatively different from that seen and expected for cosmological clustering and thus suggests another origin. We find no clustering on these scales in the FSRQ-like sub-sample. Since the two samples are derived from the same survey and analysed using the same tools we conclude that the difference in behaviour we are seeing must arise from a real astrophysical effect. Taking the typical redshift of the BL Lacs in the sample to be ∼0.35, clustering on an angular scale of 20 degrees corresponds to a linear scale of ∼350 Mpc. This is large but does not challenge the concordance cosmological model predictions that the largest structures seen should not exceed 1 Gpc.

Our results for BL Lac-like objects supports the interpretation of the large-scale alignments of quasar optical polarization position angles arise from an underlying coherent alignment of AGN axes. We remind the reader that, if what we are seeing is a result of coherent axis alignment, there does not have to be a real excess space-density of objects on this scale. This could explain why clustering analyses of AGN in general, or for that matter giant elliptical galaxies which are the hosts of BL Lacs, fail sto detect a signal on the same scale as we do. The fact that we do not detect a significant clustering signal when we examine FSRQs from the same sample gives a hint as to the maximum scale of the coherent alignments since quasars, populate on average a much larger volume than do the BL Lacs.

We suggest that the optical polarization results and what we have found might be another manifestation of the intrinsic alignment effect which is known to extend to the kind of linear scales we are talking of. We think this connection is plausible since intrinsic alignments are known to be strongest amongst red galaxies. In addition, BL Lac objects are generally hosted by red galaxies and low luminosity radio sources are the ones whose structures are known best to align with their optical hosts. All the necessary ingredients for a connection are present.

Further investigations are required to build on the present results. We are currently working on extending our analysis to radio surveys by picking flat spectrum sources most of which we expect to be blazars. Selecting suitable samples is not without problems since many radio surveys suffer from position-dependent biases which have to be recognized and eliminated as much as possible.

## ACKNOWLEDGEMENTS

We thank Lorne Whiteway for his extensive advice on different aspects of the two point correlation analysis and Neal Jackson, Scott Kay, and Richard Battye for early discussions. Peter Wilkinson and Andy Biggs provided many helpful comments on the manuscript. Alice Song helped with the implementation of the two point correlation software. M.J.M.Marchã acknowledges the support of the DiRAC-3 Operations Grant ST/S003916/1

## DATA AVAILABILITY

The data underlying this article were derived from sources in the public domain: https://fermi.gsfc.nasa.gov/ssc/data/access/lat/8yr_catalog/, and it will be shared on reasonable request to the corresponding author.